\documentclass[fleqn,usenatbib]{mnras}
\usepackage[utf8]{inputenc}
\usepackage{newtxtext,newtxmath}
\usepackage{comment}
\usepackage{multirow}
\usepackage{soul}
\usepackage[export]{adjustbox}
\usepackage[T1]{fontenc}

\DeclareRobustCommand{\VAN}[3]{#2}
\let\VANthebibliography\thebibliography
\def\thebibliography{\DeclareRobustCommand{\VAN}[3]{##3}\VANthebibliography}

\usepackage{graphicx}	
\usepackage{amsmath}	
\usepackage{amsmath}
\usepackage{geometry}
\usepackage{caption}
\usepackage{hyperref}
\usepackage{lscape}
\usepackage{amsfonts} 

\def \msun {{$M_\odot$}}

\def \flux {erg cm$^{-2}$ s$^{-1}$}

\def \xmm {{\it XMM-Newton}}
\def \src{{SAX~J1808.4--3658}}
\def \obj{{SAX~J1808}}

\def \nustar{{\it NuSTAR}}
\def \maxigsc{{\it MAXI}/GSC}
\def \nicer{{\it NICER}}

\def \relxill{{\tt relxill}}

\def \relxillCp{{\tt relxillCp}}

\newcommand{\erg}{erg cm$^{-2}$ s$^{-1}$} 
\newcommand{\lum}{erg s$^{-1}$}

\title[A broadband view at \src{}]
{Probing burst-disc interaction and disc reflection in \src{} with \nicer{}, \xmm{}, and \nustar{}}

\author[Mandal, Naik \& Chhotaray]{
Manoj Mandal\thanks{E-mail: manojmandal213@gmail.com},
Sachindra Naik, and Birendra Chhotaray \\
Astronomy and Astrophysics Division, Physical Research Laboratory, Navrangpura, Ahmedabad - 380009, Gujarat, India
}
 
\date{Accepted XXX. Received YYY; in original form ZZZ}

\begin{document}
\label{firstpage}
\pagerange{\pageref{firstpage}--\pageref{lastpage}}
\maketitle

\begin{abstract}
We performed a comprehensive study of thermonuclear bursts from the millisecond X-ray pulsar \src{} with \xmm{} and \nicer{}. We report the results from the analysis of an intense burst with \nicer{} using a self-consistent and physically motivated disc reflection modeling approach and investigate the burst-disc interaction. The dynamic evolution of the spectral parameters suggested evidence of photospheric radius expansion (PRE) of the neutron star using the disc reflection modeling approach, which indicates a maximum expansion of the photosphere up to $14.8\pm0.7$ km. The corresponding blackbody temperature drops to a minimum of 1.9 keV. In addition, an emission line at 1 keV is observed, possibly originating from the Ne or Fe L-band transition as a result of the reprocessing of burst photons by cold gas in the accretion disc. The 1 keV emission line flux is found to be strongly correlated with the flux of the reflection component. We also investigated a thermonuclear burst observed with \xmm{} EPIC-PN from \src~ using the variable persistent emission method and the disc reflection modeling approach. The X-ray reflection feature is also investigated in persistent emission using a \nustar{} observation. The best-fitting results provide an inner disc radius of $14_{-5.9}^{+9.7}$ $R_g$ and an inclination of $ 38^\circ-60^\circ$ during the \nustar{} observation. The magnetic field is estimated to be $\simeq$3.7 $\times$10$^8$ G at the poles of the neutron star.
\end{abstract}
\begin{keywords}
accretion, accretion discs -- stars: neutron – X-ray: binaries -- X-rays: bursts – X-rays: individual: \src{}
\end{keywords}
\section{Introduction}
\label{intro}
A neutron star low-mass X-ray binary (LMXB) system consists of a neutron star (NS) and a low-mass companion star ($M\leq 1\,M_\odot$) rotating around the common center of mass. Accretion occurs via the Roche-lobe overflow mechanism in the LMXB systems. In LMXBs with a low magnetic field NS ($\sim$10$^{7-9}$ G; \citealt{Ca09, Mu15}), the accreted matter directly accumulates on the NS surface. In this scenario, the accretion disc extends close to the NS surface, unlike the case of the high magnetic field NS LMXB systems ($\sim$10$^{12}$ G; \citealt{Staubert2019}), where accretion takes place at the poles. In the case of the low magnetic field NS, the accumulated matter on the surface of the NS triggers unstable burning of hydrogen and/or helium, producing thermonuclear X-ray bursts \citep{Le93, Ga06}. 

An accreting millisecond X-ray pulsar (AMXP) is a subclass of NS LMXBs, which shows a coherent X-ray pulsation of a few milliseconds (for review \citealt{Salvo2022}). The AMXP also exhibits thermonuclear X-ray bursts. Thermonuclear bursts exhibit a sharp rise in X-ray intensity for a few seconds, followed by a slow exponential decay that lasts several tens of seconds. In this process, lighter elements are burned into heavier elements via a nuclear chain reaction \citep{Le93, St03, Ga21}. In some cases, the burst peak flux reaches the Eddington limit, and the high radiation pressure lifts the photosphere above the surface of the NS, resulting in a photosphere radius expansion (PRE) event \citep{Le93, Ta84, Ga21}. At a constant luminosity level (Eddington luminosity), the blackbody temperature reaches a minimum value, and the photosphere radius gets extended to its maximum. At the end of the process, the photosphere radius is re-established to the initial value after the touchdown phase, which is indicated by the second peak in the temperature profile. The PRE burst can be used to constrain the compactness of the NS, which may help to understand the composition of the interior using the equation of state of the NS. 

The burst spectra are typically described by a blackbody model \citep{Va78, Ku03}, which assumes that the NS emits like a perfect blackbody. However, in an intense X-ray burst, the burst emission deviates from a perfect blackbody model, and a soft and hard excess emission is observed in the burst spectrum \citep{Worpel2013} as an observational consequence. The excess emission is possibly due to the Poynting-Robertson drag, which results in enhanced persistent emission \citep{Walker1992, Worpel2013}. Alternatively, excess emission can also be explained using the self-consistent and physically motivated disc reflection modeling approach that indicates the reflection of burst photons from the accretion disc \citep{Zh22, Yu24, 4U1702}. 

Excess emissions are particularly prominent during superbursts, which are highly energetic thermonuclear X-ray bursts from accreting neutron stars, likely powered by unstable carbon burning at much higher depths than a typical Type-I burst, and last for several hours while releasing energy of orders of magnitude higher than ordinary Type-I bursts, thus providing a unique probe of burst–disc interactions \citep{Keek2008}. Several observational consequences of the reflection process, such as emission and absorption lines in burst spectra, are observed in superbursts, strongly suggesting reflection from the photoionized disc \citep{Ke17, De18, Ba04}. Further soft X-ray investigations of thermonuclear bursts exhibiting the PRE events are relevant to understanding the emission mechanism and burst-disc interactions.   

 In several NS LMXBs, X-ray reflection features are observed in persistent emission \citep{Lu19, Ludlam2024, SRGA25}. Reflection spectroscopy is useful for probing the geometry and properties of the accretion disc. The magnetic field of the NS can be estimated by X-ray reflection spectroscopy, assuming a disc truncation close to the magnetosphere boundary \citep{Ca09, Lu19}. Fluorescent emission lines, including a broad iron line at 6.40-6.97 keV, and a Compton hump in the 20–40 keV range, are produced as the hard Comptonized photons scatter off the accretion disc that typically appears as a reflection spectrum. The broadening of the iron line is caused by relativistic effects close to the neutron star, Doppler effects, and scattering in the inner accretion disc \citep{Ge91, Fa89, Fa00}. 

The burst oscillation in a few hundred Hz is detected during certain thermonuclear X-ray bursts in several LMXBs. The burst oscillation during the X-ray burst is understood to be due to the asymmetric brightness patches in the burning surface layer of the NS, which can be utilized to constrain the spin, radius, and mass of the NS \citep{Watts2012}. This phenomenon is used to understand the dense matter physics at high magnetic field environments \citep{Strohmayer2006, Watts2012}. 

The NS LMXB \src{} (\obj{} hereafter) was discovered in 1996 by the {\it BeppoSAX} satellite \citep{int1998}. A coherent pulsation of $\sim$2.5 ms was reported from the source \citep{Wijnands1998}, which confirmed the source as an accreting millisecond X-ray pulsar (AMXP), the first discovered AMXP. The AMXP orbits a companion star of 0.05 \msun{} with an orbital period of $\sim$2 hr, and the source distance is approximately $\sim$3.5 kpc \citep{Ga06}. The source has undergone several outbursts since its discovery, with a recurrence time of approximately 2 to 4 years. \obj{} showed several thermonuclear X-ray bursts \citep{int1998, Ga06, Ga08} and burst oscillations at the coherent spin frequency of $\sim$401 Hz \citep{Chakrabarty2003, Bult2019}. \obj{} has exhibited prominent kilohertz quasi-periodic oscillations (kHz QPOs) during outbursts, including episodes where twin kHz QPOs were detected. These QPOs are observed as narrow features in the power density spectrum and provide evidence for accretion-disc dynamics close to the neutron star \citep{Wijnands2003}.  

In the present work, we carried out a detailed time-resolved spectroscopy of the thermonuclear bursts and also the pre-burst persistent emission in \obj. We revisited a \nicer{} observation from August 2019, during which the brightest thermonuclear burst to date was detected. \citet{Bult2019} analyzed the data during this \nicer{} X-ray burst and reported the results from spectral fitting using a double blackbody model; however, the excess emission can be explained alternatively using a physically motivated self-consistent disc reflection model. During such a powerful X-ray burst, it is expected that a major fraction of burst photons may interact with the accretion disc and get reflected. It is, therefore, insightful to investigate the X-ray reflection features using the disc reflection model during such an intense thermonuclear X-ray burst, to understand more about accretion physics and the burst disc interaction. In addition, we also used publicly available \xmm{} observation from September 2022 during which a thermonuclear burst was observed. A \nustar{} observation in August 2022 is used to investigate the X-ray reflection features in persistent emission and to investigate the disc geometry and its properties. The paper is organized as follows. The observations and data analysis methodology are described in Section~\ref{obs}. Section~\ref{res} summarizes the results. The discussion of the results is presented in Section~\ref{dis}. The summary and conclusions are presented in Section~\ref{con}.
\begin{table}
\centering
\caption{Summary of observations and X-ray bursts detected from \obj{} using \nicer, \xmm, and \nustar{} observations.}
\begin{tabular}{llccc} 
\hline	
Observatories	& Date of       & Observation & Exposure & No. of     \\
		      &observation    & ID           & (ks)    & bursts     \\
\hline
{\it NICER}     &21-08-2019     &2584010501     &17      &1   \\
{\it NuSTAR}    &22-08-2022     &80701312002    &107     &--  \\
{\xmm}          &09-09-2022     &0884700801	    &125     &1   \\
\hline
\label{tab:log_table_burst}
	\end{tabular}
\end{table}
\begin{figure*}
\centering
\includegraphics[width=\columnwidth]{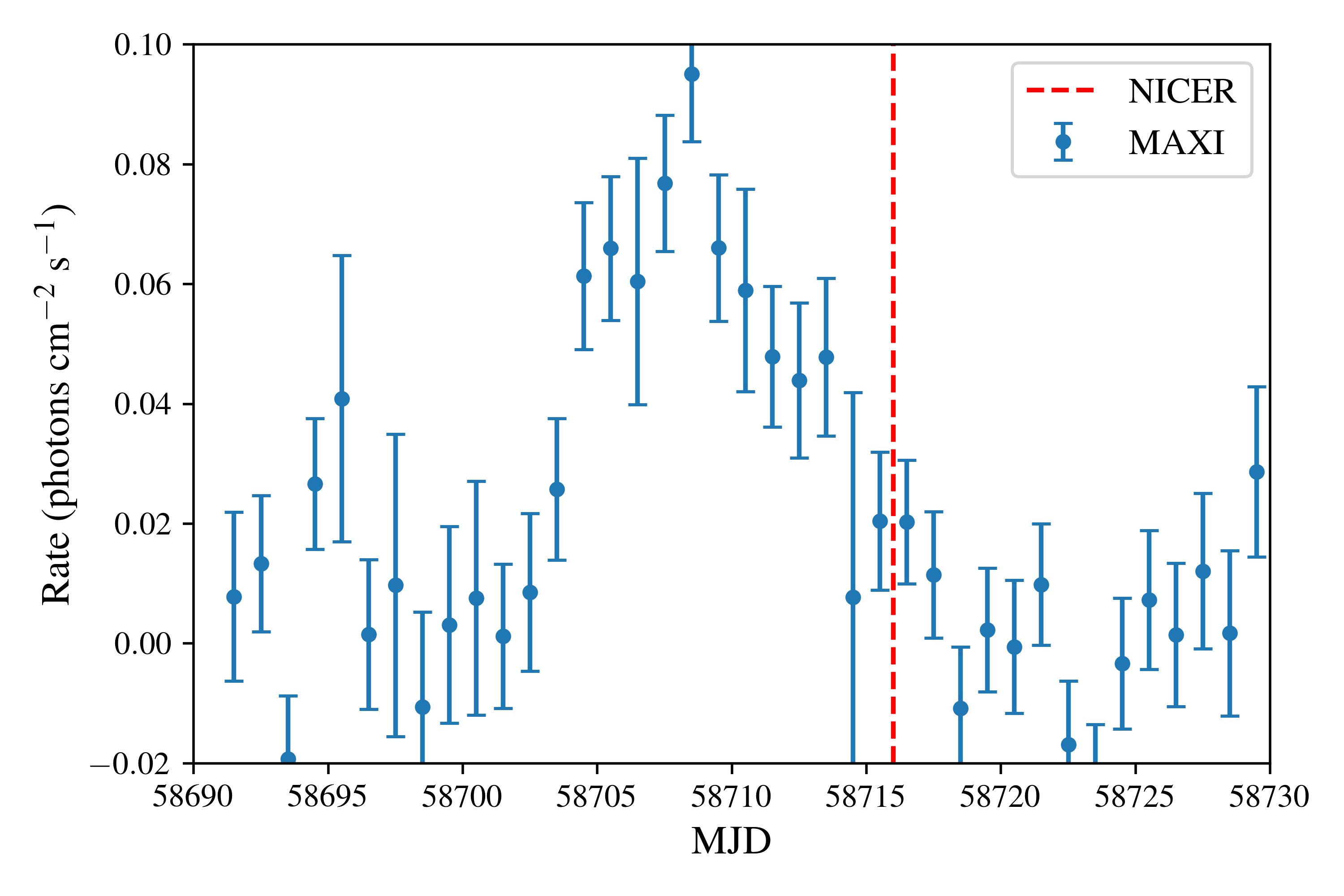}
\includegraphics[width=\columnwidth]{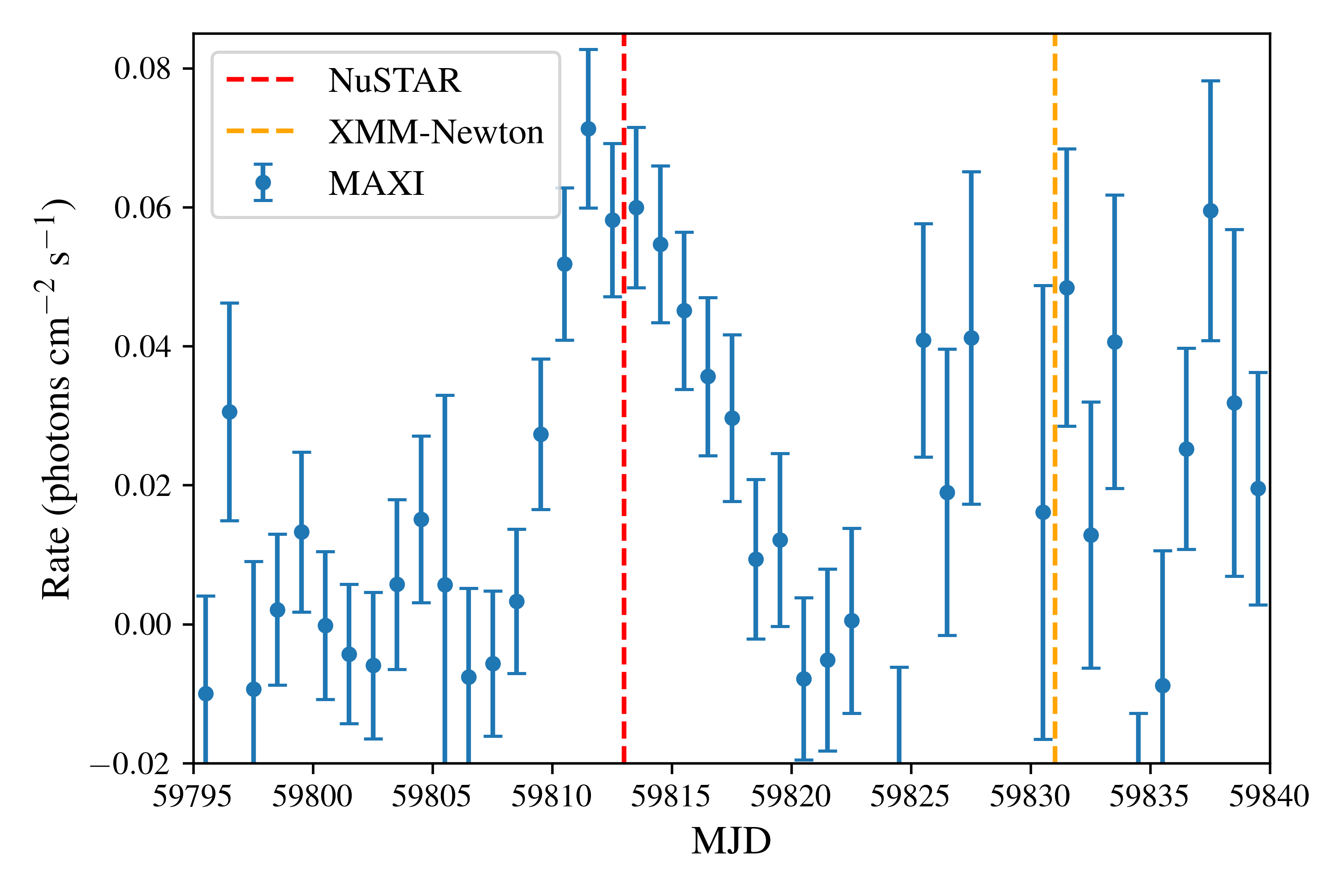}
\caption{ The figure shows the outbursts from \src\ observed in 2019 (left) and 2022 (right). Vertical dotted lines mark the observations used in this work. Thermonuclear X-ray bursts were detected during the \nicer\ and \xmm\ observations.}
    \label{fig:outburst_lc}
\end{figure*}
\section{Observation and data analysis}
\label{obs}
In this work, we used publicly available data from the observations of \obj{} with the \nicer{}, \xmm{}, and \nustar{} observatories. Data are reduced using {\tt HEASoft} version 6.33.2. 
\subsection{\nicer{} observation}
The Neutron star Interior Composition ExploreR (\nicer{}) is a non-imaging soft X-ray telescope that is operating in the energy range of 0.2--12 keV onboard the International Space Station \citep{Ge16}. \obj{} was monitored with \nicer{} during the 2019 outburst. We used a \nicer{} observation, which was performed on 21 August 2019. The details of the observation are summarized in Table~\ref{tab:log_table_burst}. The raw data are processed using the tool {\tt NICERDAS} in {\tt HEASoft} with the {\tt CALDB} version xti20240206. The {\tt nicerl2} tool is utilized to filter the raw event files. The {\tt barycorr} tool is used to perform the barycentric correction. The {\tt XSELECT} package is used to generate light curves and spectra from the barycenter corrected clean event files. The corresponding background spectrum is generated using the {\tt nibackgen3C50} tool \citep{Re22}. The response and ancillary response files are generated using {\tt nicerrmf} and {\tt nicerarf}. The spectrum is grouped for a minimum of 25 counts bin$^{-1}$ using  {\tt grppha} to apply the chi-square statistic in the analysis. 
\subsection{\xmm{} observation}
The European Photon Imaging Camera (EPIC, 0.1-15 keV) is mounted in the focus of each of the three 1500 cm$^2$ X-ray telescopes of the \xmm{} observatory \citep{Ja01}. The EPIC imaging spectrometer uses PN CCD \citep{St01}, and two MOS CCDs \citep{Tu01}. Two identical Reflection Grating Spectrometers (RGS) are placed behind the telescopes and operate in the 0.35-2.5 keV range \citep{He01}. \obj{} was observed in EPIC-PN timing mode with the \xmm{} observatory on 9 September 2022 for an exposure of $\sim$125 ks. The  \xmm{} raw data are analyzed using the \xmm{} Science Analysis System ({\tt SAS}) version 20.0.0. Events are filtered using the {\tt evselect} routine with the criteria of FLAG = 0 and PATTERN $\le4$. The {\tt epproc} is used to generate EPIC-PN event files, and the source and background light curves and energy spectrum are created using the {\tt evselect} routine. The source events are extracted from a 20-pixel-wide strip centered on the source position RAWX = 37 (i.e., RAWX in [27:47]), and the background region is selected from a source-free region in RAWX columns [3:5] from the same CCD. Spectral response files are generated using the tasks {\tt rmfgen} and {\tt arfgen}. The background correction to the light curve is performed with {\tt epiclccorr}. The EPIC-PN spectra were grouped for a minimum of 25 counts bin$^{-1}$ using  {\tt grppha} to apply the chi-square statistic in the analysis. 
\subsection{\nustar{} observation}
The Nuclear Spectroscopic Telescope Array (\nustar{}) is made up of two identical co-aligned detectors (FPMA/FPMB), which operate in the wide energy range of 3-79 keV \citep{Ha13}. In this work, we used a \nustar{} observation of \obj{} to perform a detailed spectral study of persistent emission. \nustar{} observed the source on 22 August 2022, for an exposure of $\sim$107 ks.  The details of the observation are summarized in Table~\ref{tab:log_table_burst}. To analyze \nustar{} data, we used the standard analysis software (NUSTARDAS v2.1.2) in {\tt HEASoft} v6.33.2, along with the latest {\tt caldb} version 20250428. The {\tt nupipeline} task is used to filter the event files. A circular region of 80 arcsec radius, centered on the source position, is used to extract the source events. The background region is selected away from the source. The light curves, spectra, and related response files for the source and background are generated using the task {\tt nuproducts} for both detectors. The background correction of the light curves is performed by using the {\tt lcmath} task. The spectra are grouped with {\tt grppha} with a minimum of 25 counts bin$^{-1}$.
\section{Results}
\label{res}
In this work, we performed a comprehensive study of the thermonuclear bursts detected with \nicer{} and \xmm{} from the AMXP \obj{}. The \nicer\ observation was obtained during the decay phase of the 2019 outburst, while the \xmm\ observation corresponds to the decay phase of the 2022 outburst. Figure~\ref{fig:outburst_lc} shows the \maxigsc~ 2-20 keV light curves of the 2019 (left) and 2022 (right) outbursts, with the epoch of observations used in this work marked by vertical dotted lines. The time-resolved spectral analysis is conducted to understand the dynamic evolution of spectral parameters during each burst using the variable persistent emission method and the disc reflection modeling approach. The burst observed with \nicer{} is the most intense burst to date, providing a unique opportunity to investigate the disc reflection modeling approach to understand the burst-disc interaction. In addition, an emission line at 1 keV is observed during the intense burst, which is further investigated in detail to understand the origin and emission mechanism. A \nustar{} observation of the source is also utilized to study the X-ray reflection features during persistent emission using the relativistic disc reflection model.
\subsection{Burst profile and hardness ratio}
\begin{figure}
\centering
\includegraphics[width= \columnwidth]{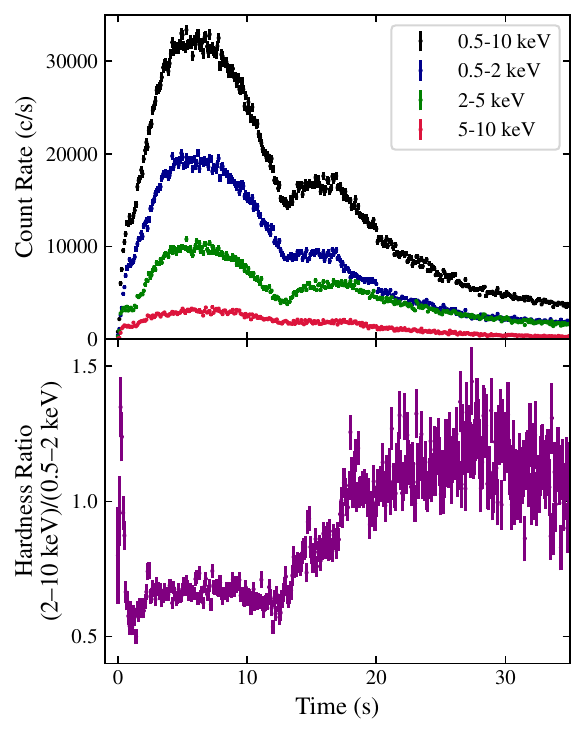}
\caption{Burst profile, along with its energy dependence, is shown in the top panel. The bottom panel represents the evolution of the hardness ratio during the thermonuclear burst observed with \nicer{}. }
    \label{fig:HR}
\end{figure}

 We investigated the most intense burst from AMXP \obj{} using a \nicer{} observation on 21 August 2019. The peak count rate reached a record high of $\sim$35000 counts s$^{-1}$ in the 0.5-10 keV range. The duration of the burst was about 30 s. To investigate the energy dependence of the burst profile, light curves in the 0.5-2 keV, 2-5 keV, and 5-10 keV ranges are created. The burst profile in different energy ranges is shown in Fig.~\ref{fig:HR}. The peak count rates in these three respective energy bands are approximately 20000 counts s$^{-1}$, 10000 counts s$^{-1}$, and 4000 counts s$^{-1}$, respectively. The burst profile shows a secondary peak-like feature in all energy bands after $\sim$15 s from the burst onset.
 
 The secondary peak-like feature is more prominent in the 2-5 keV energy band. The hardness is generated by taking a ratio between light curves in the 2-10 keV and 0.5-2 keV bands, and shown in the bottom panel of Fig.~\ref{fig:HR}. The hardness ratio shows a significant evolution during the burst. The ratio attains its minimum at the peak of the burst and remains almost constant during the flat peak, indicating that the burst peak is dominated by the photons in the 0.5-2 keV range over the hard photons. We also investigated a thermonuclear burst detected in the \xmm{} EPIC-PN observation in September 2022.  The peak count rate during this burst was $\sim$2800 counts s$^{-1}$ in the 0.5-10 keV range, which is equivalent to 3490~counts~s$^{-1}$ in the \nicer\ 0.5–10~keV band using {\tt PIMMS}\footnote{\url{https://heasarc.gsfc.nasa.gov/cgi-bin/Tools/w3pimms/w3pimms.pl}} tool. During the \xmm{} burst, the rising phase and peak were detected. However, in the decay part, $\sim$14 s of the burst were not covered during the \xmm{} observation. 
\begin{table}
\centering
 \caption{Best-fit spectral parameters [{\tt XSPEC} model : $\texttt{TBabs}\times( \texttt{po}+\texttt{diskbb})$] of the pre-TNB \nicer{} (0.5--10 keV range) and \xmm{} EPIC-PN (0.5--10 keV range) spectra of \obj{}.}
\begin{tabular}{llcc}
\hline
Components & Parameters & \nicer{} & \xmm{}  \\
& & pre-TNB & pre-TNB  \\
\hline
{\tt tbabs} & N$_{\textrm{H}}$ & $0.3\pm0.1$ & $0.1\pm0.01$  \\
\\
{\tt power-law} & $\Gamma$ & $1.7\pm0.3$ & $2.14\pm0.02$   \\
  & Norm. & 0.03$\pm$0.02 & 0.05$\pm$0.001   \\
\\
{\tt diskbb} & T$_{\textrm{in}}$(keV) & $0.7\pm0.03$ & -    \\
  & Norm. & 31$\pm$8.0 & -   \\
\\
 & Flux$_\mathrm{Total}$  & $5.01 \pm 0.05$ & $5.17 \pm 0.05$    \\
  & Luminosity  & $7.34 \pm 0.07$  & $7.57 \pm 0.07$    \\
 & $\dot{m}$  & 0.041 & 0.043      \\
 & $\frac{\dot{m}}{\dot{m}_{\mathrm{Edd}}}$  & 0.0047 & 0.0049       \\
\\
 & $\chi^{2}$/dof & 295/325 & 713/667     \\
\hline
\label{tab:tab_preburst}
\end{tabular}\\
Note: The reported errors are of 90\% significance. All fluxes (0.1--100 keV range) are corrected for absorption and quoted in units of 10$^{-10}$ ergs cm$^{-2}$ s$^{-1}$. The X-ray luminosity is in the unit of 10$^{35}$ erg s$^{-1}$ (assuming a source distance of 3.5 kpc), $N_\textrm{H}$ is in the unit of $\rm 10^{22}~ cm^{-2}$. The mass accretion rates $\dot{m}$, $\dot{m}_{\mathrm{Edd}}$ are in units of 10$^{4}$~g~cm$^{-2}$ s$^{-1}$, The Eddington rate ($\dot{m}_{\mathrm{Edd}}$) for a typical NS of radius 10 km is assumed to be $8.8\times10^{4}$ g cm$^{-2}$ s$^{-1}$ \citep{Ga08}. 
\end{table}
\subsection{Time-resolved burst spectroscopy}
\label{time_resolved_spectroscopy}
Time-resolved spectroscopy is performed to investigate the dynamic evolution of the spectral parameters for each thermonuclear burst (TNB). In this work, we conducted a detailed time-resolved spectral analysis of the burst observed with \nicer{} and \xmm{} EPIC-PN.  The \nicer~ observation was in the decay phase of the 2019 outburst, for which the type-I X-ray burst was detected. The \xmm~ observation was performed during the 2022 outburst (see Fig. \ref{fig:outburst_lc}). The contribution of persistent emission is also taken into account by analyzing the pre-TNB persistent spectrum of each burst. Persistent spectra are generated for an exposure of $\sim$1 ks before the onset of the bursts. The \nicer{} persistent spectrum is modeled using an absorbed power law and a disc blackbody model [{\tt {XSPEC: tbabs $\times$ (power law + diskbb)}}]. The best-fitted spectral parameters are presented in Table~\ref{tab:tab_preburst}. The value of equivalent hydrogen column density is found to be $\sim0.3 \times 10^{22}$ cm$^{-2}$ and the photon index and the blackbody temperature are 1.7 and 0.7 keV, respectively. The persistent emission spectrum from \xmm{} EPIC-PN can be well fitted with an absorbed power-law model, yielding a photon index of 2.1. The best-fit model provides a reduced $\chi^{2}$ of $\sim1.0$ in both cases. The unabsorbed flux (0.1-100 keV) is estimated to be in the range (5.0-5.2) $\times 10^{-10}$ erg cm$^{-2}$ s$^{-1}$ for the \nicer{} and \xmm{} observations. These best-fit spectral parameters are used during the modeling of the time-resolved spectral study of each thermonuclear burst.
\begin{figure}
\centering
     \includegraphics[width=0.70\columnwidth, angle=270]{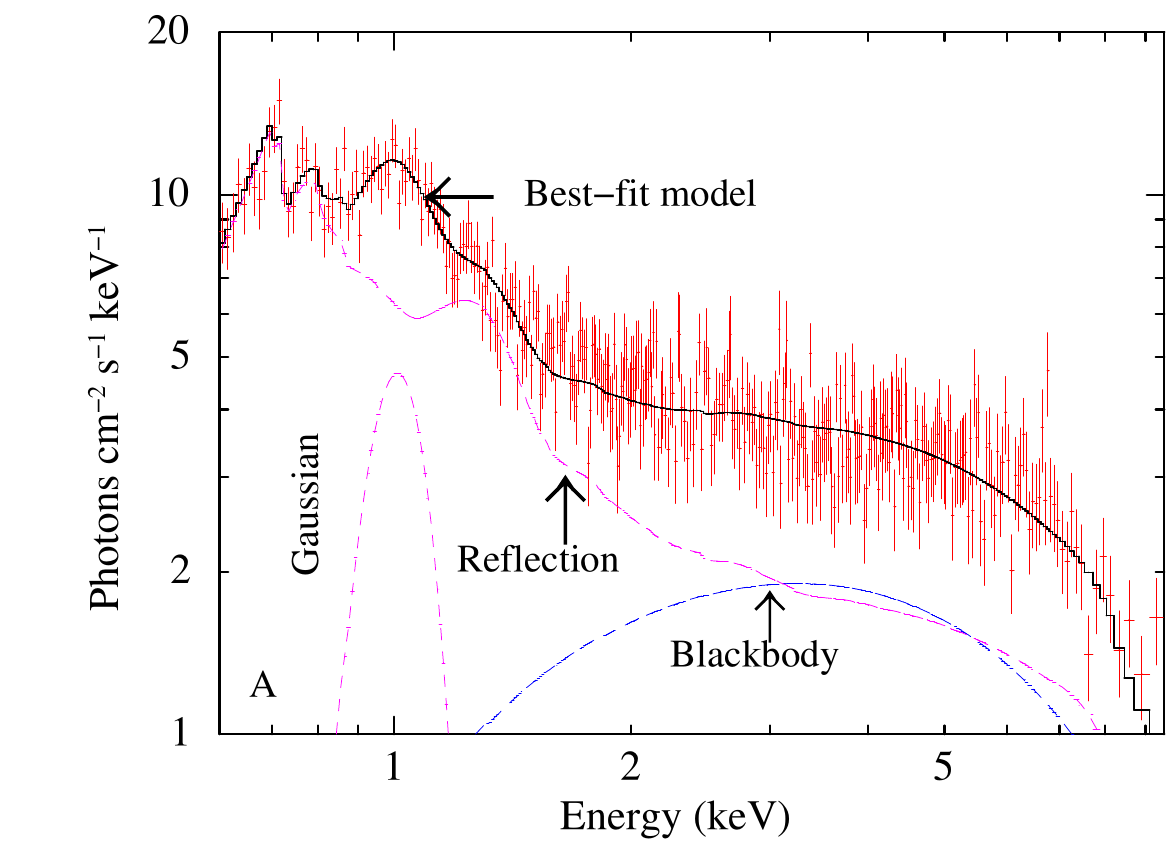} 
     \includegraphics[width=0.63\columnwidth, angle=270]{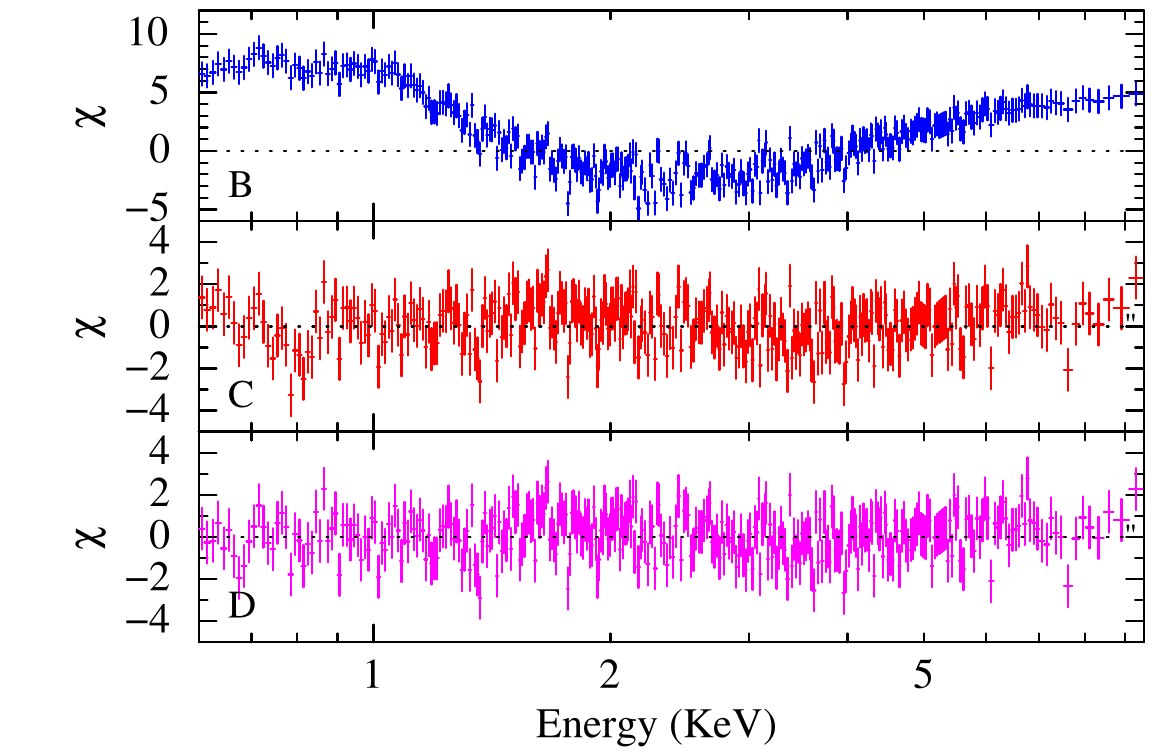}
\caption{ The best-fit spectrum of \obj{} at the peak of the thermonuclear burst observed with \nicer, fitted with an absorbed blackbody, a disk reflection model, a Gaussian component at 1 keV, and an absorption edge at 0.7 keV, is shown along with the spectral components (panel~A). The pronounced positive residuals at both ends of the energy band, when the spectrum is fitted with a blackbody model alone, are shown in panel~B. Adding a reflection component significantly improves the fit and removes most of these residuals, while a Gaussian component accounts for the 1 keV emission feature (panel~C). The best-fit residuals are obtained after including an absorption edge at 0.7 keV and are shown in the bottom panel (panel~D).}
\label{fig:reflection_nicer}
\end{figure}

We perform a detailed study of the most intense \nicer{} burst from \obj{}. To perform the time-resolved spectral analysis, the burst is divided into 26 segments based on count statistics. During the peak, a comparatively smaller bin time of 0.5 s is used, and the bin time is gradually increased during the decay part to optimize the signal-to-noise ratio. Spectra and related responses are generated for each segment by applying suitable good time intervals (GTI). The background spectra are generated using the {\tt nibackgen3C50} tool. Initially, the \nicer{} burst spectra are modeled with an absorbed blackbody model [{\tt {XSPEC: tbabs $\times$ bbodyrad}}] and the contribution from persistent emission is accounted for by using the best-fit pre-burst spectral parameter values. However, the spectra are not well fitted by this model and show significant residuals/excess emission at both ends of \nicer{} energy bands. The residuals of the spectra fitted with an absorbed blackbody model are shown in panel~B of Fig.~\ref{fig:reflection_nicer}. The residuals show an excess emission below 2 keV and above 6 keV. To account for the excess emission, we first followed the variable persistent emission approach ($f_a$ method), in which the excess emission is explained in terms of the enhanced accretion rate of persistent emission \citep{Worpel2013}. The modeling using the $f_a$ method yields a $f_a$ value exceeding 150 during the burst peak, which is unrealistic and unacceptable at the peak of the burst. The $f_a$ modeling approach provides a blackbody temperature of $\sim$3.2 keV and $f_a$ parameter of $\sim$180. The fitting results provide a statistically unacceptable fit with $\chi^2$/dof = 935/342. This model does not fit the large residual below 1.8 keV. The obtained scaling factor ($f_a$) is unrealistically large, exceeding the commonly observed range of $f_a$ = 2–10 \citep{Worpel2013}. For these reasons, we do not pursue the $f_a$ model as an explanation for the observed spectra. However, the $f_a$ modeling approach does not provide any physically motivated explanation of the excess emission. 
\begin{figure}
    \includegraphics[width=0.95\columnwidth]{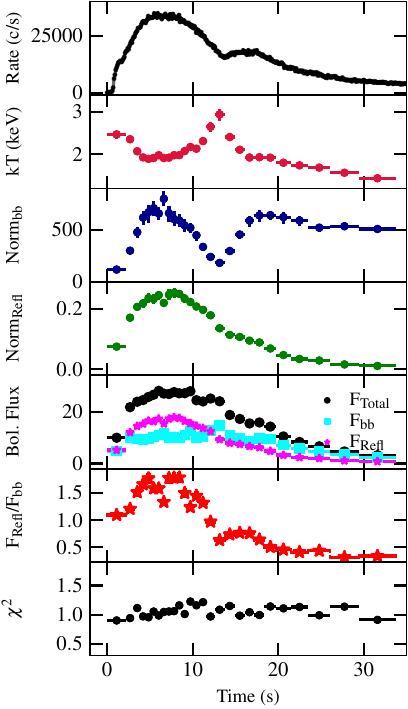} 
\caption{Evolution of spectral parameters of \obj{} during the thermonuclear burst observed with \nicer, using an absorbed blackbody and a relativistic disc reflection model {\tt relxillNS}. In addition, a Gaussian at 1 keV and an edge at 0.7 keV are also included in the model.}
    \label{fig:fig_time_Resolved_nicer}
\end{figure}

During the intense burst, a fraction of burst photons may interact with the accretion disc and be reflected, which strongly motivated us to investigate the excess emission with a disc reflection modeling approach and explore the burst-disc interaction. We used the relativistic disc reflection model {\tt relxillNS} \citep{Garcia2022} to explain the excess emission. This provides a self-consistent and physically motivated explanation. The {\tt relxillNS} model assumes that the photoionized accretion disc is illuminated by a blackbody spectrum from the NS. The disc reflection model {\tt relxillNS} includes the model parameters such as: inner and outer disc radii $R_\mathrm{in}$, $R_\mathrm{out}$, disc inclination $i$,  dimensionless spin parameter $a$, inner and outer emissivity indices $q1$ and $q2$, iron abundance $A_\mathrm{Fe}$, disc ionization parameter $\log \xi$, disc density $\log \text{N}$, and input blackbody temperature $kT_\mathrm{bb}$. It is challenging to constrain every model parameter at once in such short time intervals. The disc parameters in the {\tt relxillNS} model are adopted based on previous studies of similar neutron-star low-mass X-ray binaries \citep{Lu19, Zh22, Lu23, Yu24,Ma25, 4U1702}. Earlier works have shown that the accretion discs in such systems become dense and highly ionized, particularly during thermonuclear bursts, with typical densities of $\log N \sim 10^{18-20}$ cm$^{-3}$ \citep{Gu22, Ja24} and ionization parameters $\log \xi \gtrsim 3.0$ erg cm s$^{-1}$ \citep{Lu19, Zh22, Ja24}. The iron abundance was initially allowed to vary in a few time segments during the burst. Since it provides a value of $A_{\rm Fe} \sim 5$ for most of these segments, we fixed it at $A_{\rm Fe} = 5$ for the remaining time-resolved fits to improve stability and avoid degeneracy, following earlier {\tt relxillNS} burst studies. A similar high iron abundance was reported during such intense bursts \citep{Zh22, Yu24, Ma25}.
Therefore, the following parameters are frozen at the typical values for similar systems during spectral modeling: assuming a single emissivity profile $q1 = q2$ = 3, $A_\mathrm{Fe}$ = 5, $i$ = 60$^\circ{}$, ${\log N}$ = 19 cm$^{-3}$, $R_\mathrm{in} = R_\mathrm{ISCO}$, $R_\mathrm{out}$ = 400 $R_g$, and $\log \xi$ = 3.1 erg cm s$^{-1}$. For the {\tt relxillNS} model, the reflection fraction parameter is set to $-1$, where the negative sign accounts for only the reflection emission \citep{Zh22, Yu24, Ma25}. The temperature of the input blackbody spectrum in {\tt relxillNS} is tied to the temperature of the {\tt bbodyrad} component. This means that the {\tt bbodyrad} model describes the direct coronal emission, while the {\tt relxillNS} describes the reflection component. The normalization of the reflection model component is allowed to vary during time-resolved spectral modeling. The burst spectra are well described by the model consisting of absorbed blackbody and reflection model components [{\tt XSPEC: tbabs $\times$ (bbodyrad + relxillNS)}], while the pre-burst persistent emission is accounted for using the best-fit pre-burst parameters summarized in Table~\ref{tab:tab_preburst}. Though most of the features are well modeled with this model combination, an additional emission line at $\sim 1$ keV is evident in the residuals of the burst spectra. Therefore, an additional Gaussian component is included in the model. The residuals obtained by fitting this model to the spectrum are given in panel~C of Fig.~\ref{fig:reflection_nicer}. However, this provides an acceptable fit, with residuals around 0.7-0.8 keV. We therefore added an absorption edge to model this feature. The residuals after including the edge component are shown in panel D of Fig.~\ref{fig:reflection_nicer}. 

In addition, as the {\tt relxillNS} model contains the incident blackbody illuminating continuum, we examined fitting of all the burst segments with the {\tt relxillNS} model by making the reflection fraction ($R_{\rm f}$) a free positive parameter. This model setup yields results similar to those obtained with the earlier configuration, with the reflection fraction reaching a maximum value of $\sim2.3$. The evolution of the input continuum temperature ($kT$), the reflection fraction ($R_{\rm f}$), total observed flux (sum of illuminating continuum and reflected flux), and reduced $\chi^2$ are investigated across the burst duration. We find that the above parameters are consistent within uncertainties with those parameters obtained from the model consisting of a blackbody component and {\tt relxillNS} with $R_{\rm f}$ fixed at $-1$. As the {\tt relxillNS} model with a free positive $R_{\rm f}$ does not provide the normalization of the built-in illuminating blackbody component, required to estimate the radius of the emitting region, we adopt {\tt bbodyrad + relxillNS} model (with $R_{\rm f}=-1$) in our spectral fitting.

The dominance of the reflected flux over the illuminating blackbody flux (R$_f$ > 1) during the PRE phase allowed us to explore the contribution of emission from the accretion disk towards the total emission. For this, we included an additional {\tt diskbb} component to the {\tt bbodyrad + relxillNS} (with $R_{\rm f}=-1$) model to account for the excess in low-energy X-rays and estimate its contribution towards the overall energy budget. However, the addition of {\tt diskbb} component does not improve the fits for the majority of the segments. The values of some spectral parameters are poorly constrained, whereas the values of some parameters are not physically meaningful throughout the burst duration; e.g., the inner disk radius is derived to be in the range 1.6--200 km. Moreover, the inferred {\tt diskbb} flux is about two orders of magnitude lower than the flux of the blackbody and reflection components. Therefore, we adopt the {\tt bbodyrad + relxillNS} model as our preferred spectral description.
\begin{figure}
\centering
    \includegraphics[width=0.8\columnwidth]{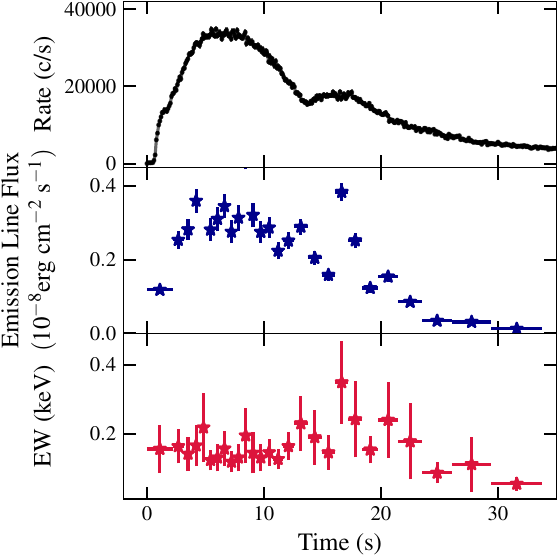}    
\caption{Evolution of equivalent width (EW) and flux of the emission line at 1 keV in the spectrum of \obj{} during the thermonuclear burst observed with \nicer.}
    \label{fig:fig_time_Resolved_nicer2}
\end{figure}

As described above, we carried out our time-resolved spectral analysis by using the {\tt bbodyrad + relxillNS} model for the \nicer~ burst. The evolution of the blackbody temperature and the corresponding normalization indicate a PRE event during the thermonuclear burst observed with \nicer. During the PRE event, the blackbody temperature reached a minimum value of $\sim$1.9 keV, and the corresponding normalization of the blackbody model was $799_{-71}^{+73}$. The {\tt bbodyrad} model normalization is reported in standard XSPEC units $(R_{km}/D_{10})^2$. The radius of the apparent blackbody-emitting region is estimated to be $9.9\pm0.4$ km, assuming a source distance of 3.5 kpc. The peak flux of the blackbody and the reflection components are $(14.7\pm0.5) \times 10^{-8}$ \flux{} and $(17.8\pm0.3) \times 10^{-8}$ \flux{}, respectively. The total bolometric flux reached a highest value of $(27.9\pm0.3) \times 10^{-8}$ \flux{} at the peak of the burst. An additional absorption edge was included to model the feature near 0.7~keV. The edge energy is $E = 0.72^{+0.02}_{-0.03}$~keV and the optical depth is  $\tau = 0.40^{+0.16}_{-0.13}$ (90\% confidence). Inclusion of this component improves the fitting results from  $\chi^2/\mathrm{dof} = 371/339$ to $352/337$. The best-fitted spectrum during the burst peak, along with the model components, is shown in the top panel of Fig.~\ref{fig:reflection_nicer} (panel~A) and the corresponding residuals are presented in the bottom panel (panel~D) of the figure.

\begin{figure*}
    \includegraphics[width=0.65\columnwidth]{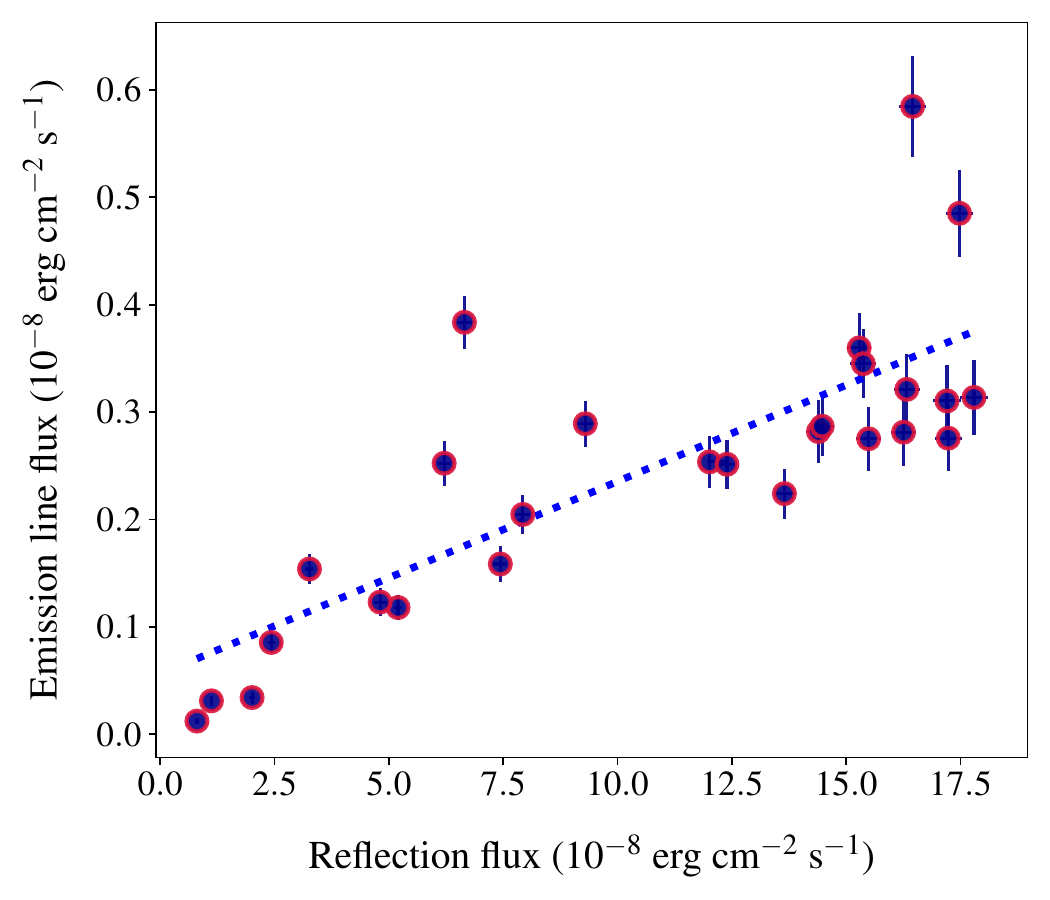}   
    \hspace{0.01\textwidth}%
     \includegraphics[width=0.65\columnwidth]{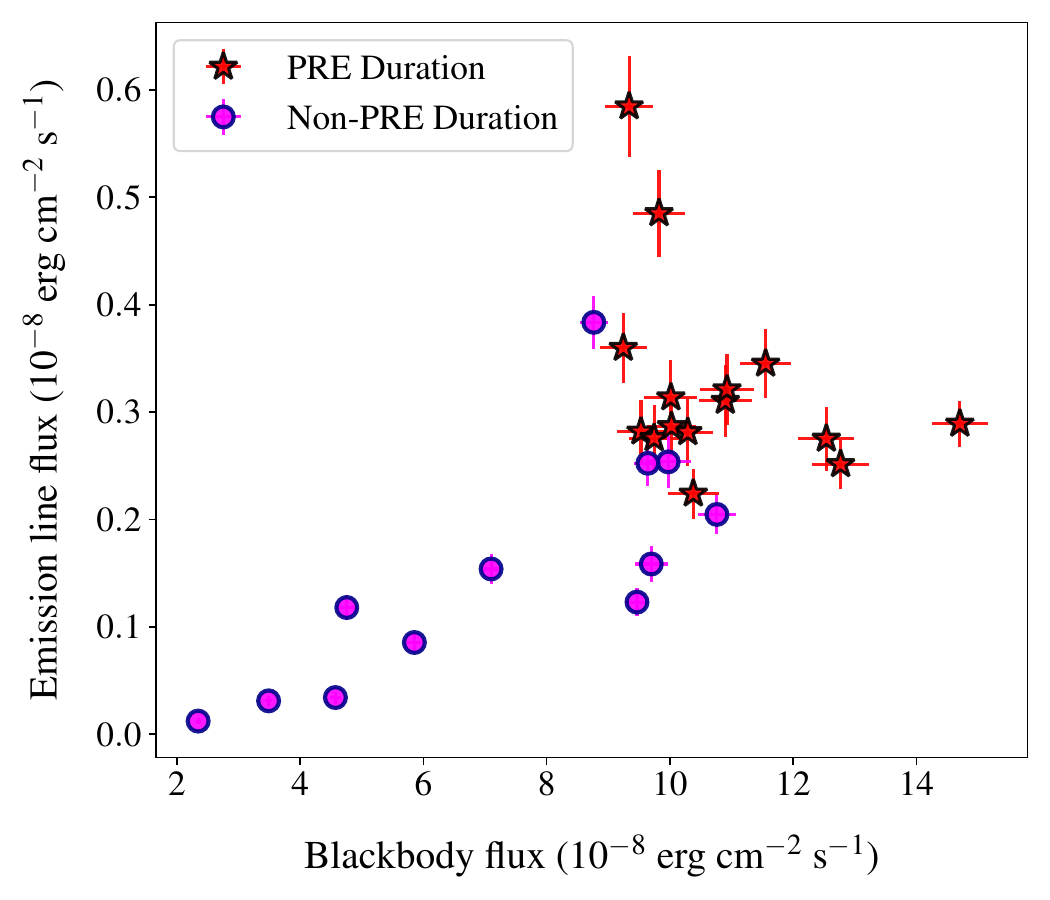} 
       \hspace{0.01\textwidth}%
      \includegraphics[width=0.65\columnwidth]{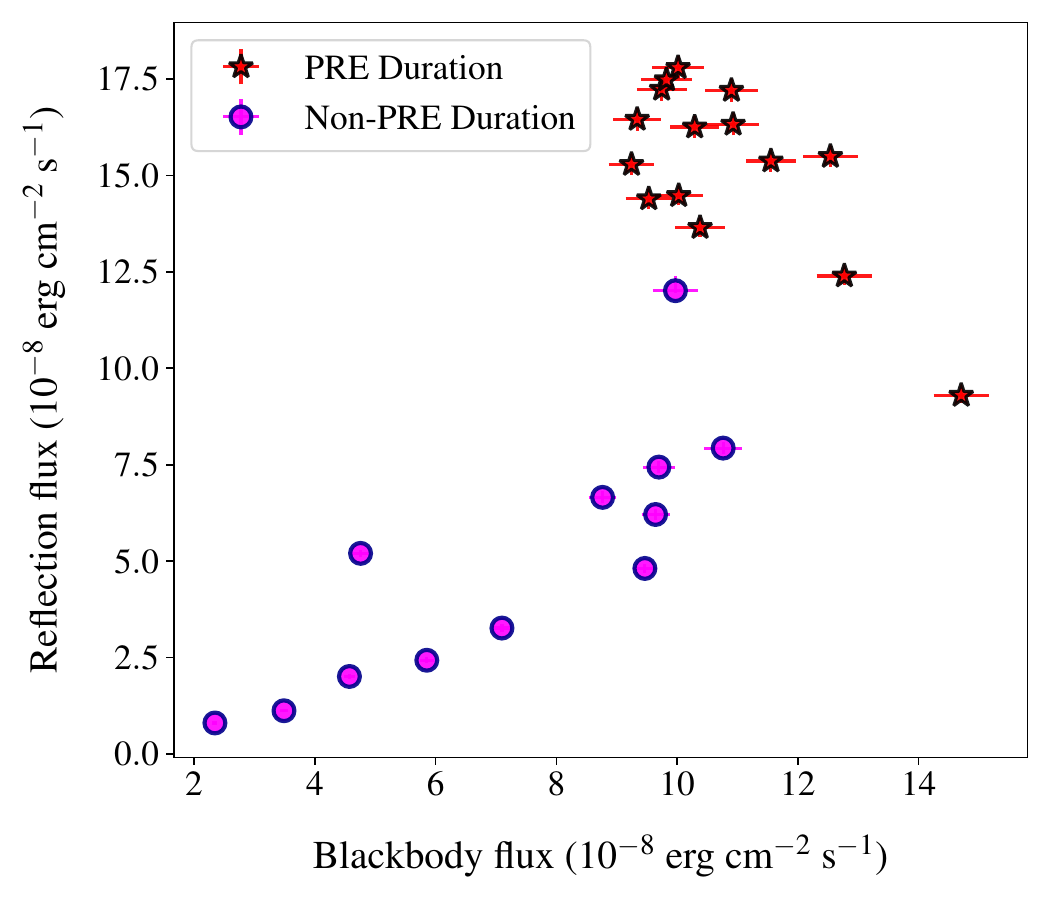} 
\caption{Correlations between the 1 keV emission line flux and the reflection component flux (left panel), the 1 keV emission line flux and blackbody flux (middle panel), and the reflection component flux and blackbody flux (right panel) are shown. The dotted line in the left panel shows the best-fit linear regression fit.  The Pearson correlation coefficient and the p-value are 0.79 and $10^{-7}$, respectively.  The linear regression analysis yields a coefficient of determination $R^2 = 0.63$, indicating that 63\% of the variance in the emission-line flux is explained by the reflection flux.}
    \label{fig:correlation2}
\end{figure*}

The evolution of the burst spectral parameters, including the blackbody temperature (kT in keV), normalization of the blackbody component, normalization of the reflection component, and bolometric fluxes of different model components, is shown in Fig.~\ref{fig:fig_time_Resolved_nicer}. In addition, the ratio of the reflection flux to the blackbody flux (F$_\mathrm{Refl}$/F$_\mathrm{bb}$) over the 0.1-100 keV range is shown in the sixth panel of Fig.~\ref{fig:fig_time_Resolved_nicer}. This ratio reaches a maximum of $\sim$1.7 and remains above unity during only the PRE phase, indicating the presence of strong reflection during this phase. The evolution of the 1 keV emission line flux and the equivalent width of the line are shown in the middle and bottom panels of Fig.~\ref{fig:fig_time_Resolved_nicer2}, respectively, along with the burst light curve in the top panel for comparison. The figure shows that the emission line flux followed the burst light curve, whereas the equivalent width of the emission line varied between 60 and 300 eV. The equivalent width remained nearly in the range of 110-220 eV during the peak and decreased to $\sim$60 eV during the cooling tail. The correlation between the 1 keV emission line flux and the flux of the reflection and blackbody components is examined and shown in Fig.~\ref{fig:correlation2}. The emission line flux shows a strong linear correlation with the flux of the reflection component (left panel of Fig.~\ref{fig:correlation2}). To verify the strength of the correlation, the Pearson correlation test is performed, yielding a correlation coefficient of 0.79 and a p-value of 10$^{-7}$. The correlations between the emission line flux and the flux of the blackbody component, and blackbody flux and reflection flux are shown in the middle and right panels of Fig.~\ref{fig:correlation2}, respectively. The points corresponding to the PRE phase of the burst are shown in a different color (red asterisk). Though the emission line flux and blackbody flux (middle panel), and blackbody flux and reflection flux (right panel) show positive correlations, these are deviated from the strong linear correlation seen in the case of emission line flux and reflection flux. This deviation during the PRE phase is due to the blackbody flux being nearly constant, whereas the reflection/emission line flux shows an increasing trend. 

\begin{figure*}
    \includegraphics[width=0.85\columnwidth]{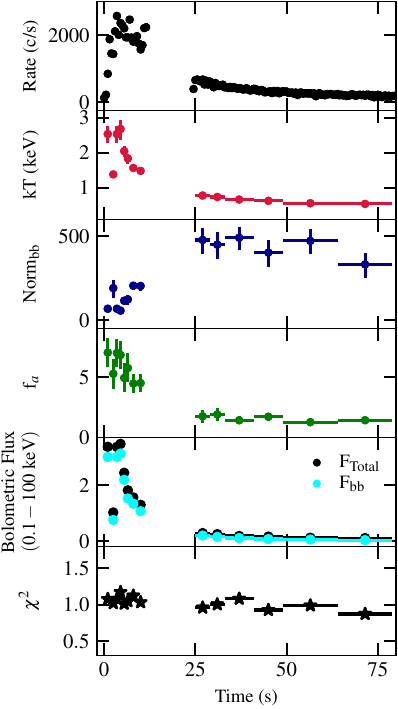} 
    \hspace{0.035\textwidth}%
    \includegraphics[width=0.85\columnwidth]{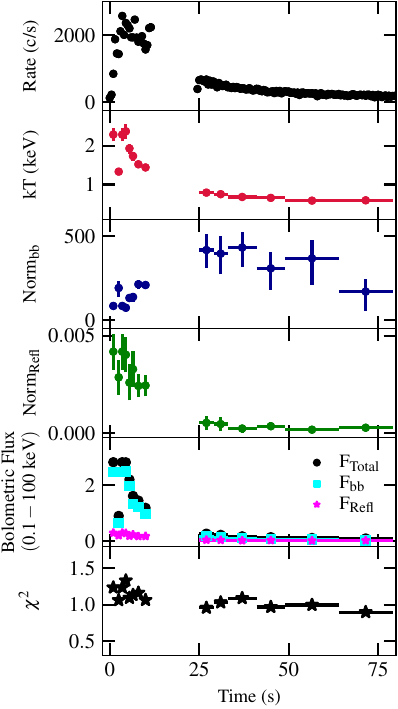}  
\caption{The variation of several spectral parameters of \obj{} during the thermonuclear burst observed with \xmm{} EPIC-PN. The time-resolved burst spectra are modeled using the variable persistent emission method (left) and disc reflection model (right). The reported flux values are in the unit of 10$^{-8}$ \flux{}.}
    \label{fig:fig_time_Resolved_xmm}
\end{figure*}

We also performed a time-resolved spectral analysis of the X-ray burst from \obj{} detected with \xmm{} EPIC-PN. During this burst, unlike the one observed with \nicer{}, there is no signature of the PRE event. The \xmm{} EPIC-PN burst is divided into fourteen segments to investigate the evolution of the spectral parameters. The time-resolved burst spectra are modeled using the absorbed blackbody model [{\tt XSPEC: tbabs $\times$ bbodyrad}] and persistent emission is taken into account by freezing the preburst model parameters at best-fit preburst values (see Table~\ref{tab:tab_preburst}). The burst spectra deviate from pure blackbody emission due to the soft excess emission below 2 keV, as seen in the residuals. The excess emission is further explained through two different approaches: the variable persistent emission method ($f_a$ method) and the relativistic disc reflection modeling approach. Fig.~\ref{fig:fig_time_Resolved_xmm} (left side) shows the evolution of spectral parameters during the thermonuclear burst, including the blackbody temperature (second panel), normalization (third panel), and bolometric flux (fifth panel). The $f_a$ parameter, which is varied between 1.5 and 7, reached a maximum value of $\sim$7 during the peak of the X-ray burst (fourth panel), the blackbody temperature reached a peak value of $\sim$2.5 keV, and the corresponding blackbody flux reached a value of $\sim$3 $\times$ 10$^{-8}$ erg cm$^{-2}$ cm$^{-1}$. The burst light curves are shown in the upper panels of Fig.~\ref{fig:fig_time_Resolved_xmm} to visualize the evolution of spectral parameters during the burst. The reduced $\chi^{2}$ values for the time-resolved spectra are shown in the bottom panel of Fig.~\ref{fig:fig_time_Resolved_xmm}. The evolution of spectral parameters using the disc reflection model is shown in the right panel of Fig.~\ref{fig:fig_time_Resolved_xmm}. The blackbody temperature and normalization show a trend similar to the $f_a$ method. In addition, the normalization of the reflection model is shown in the fourth panel in Fig.~\ref{fig:fig_time_Resolved_xmm}.

Alternatively, excess emission can be modeled using the disc reflection model {\tt relxillNS}. The model parameters assumption is similar to the \nicer{} burst spectral modeling approach with the disc reflection model. The normalization of the disc reflection model is allowed to vary during the burst, and the disc parameters are set to the typical values of a similar system. The evolution of spectral parameters, including the normalization of the reflection model and the flux of the reflection component, is shown in the right panels of Fig.~\ref{fig:fig_time_Resolved_xmm}. The top panel of Fig.~\ref{fig:fig_time_Resolved_xmm} shows the burst light curve of \obj{} observed with \xmm{} EPIC-PN. The blackbody temperature, normalization of the blackbody component, normalization of the reflection component, and bolometric flux are shown in the second, third, fourth, and fifth panels of Fig.~\ref{fig:fig_time_Resolved_xmm}, respectively. The blackbody temperature reached a maximum of 2.2 keV, and the corresponding blackbody flux is estimated to be $\sim$2.5 $\times$ 10$^{-8}$ erg cm$^{-2}$ cm$^{-1}$. 
\subsection{X-ray reflection from persistent emission}
\label{reflection_persistent}
The X-ray reflection feature is investigated in the persistent emission of \obj{} using a \nustar{} observation in September 2022, owing to its broadband coverage. The persistent \nustar{} spectra are modeled using a single-component continuum model, such as the power-law modified by the {\tt TBabs} component, the single temperature blackbody component ({\tt bbodyrad} in {\tt XSPEC}), the thermal Comptonization component ({\tt nthComp} in XSPEC: \citep{Zd96, Zy99}), and the multi-temperature disc blackbody component ({\tt diskbb} in {\tt XSPEC}: \citet{Mi84, Ma86}). The line-of-sight absorption due to Galactic neutral hydrogen is considered with the {\tt TBabs} model \citep{Wilms2000} in the {\tt XSPEC}. The most recent abundance model, {\tt wilm}, is adopted and incorporated into {\tt XSPEC} for the input abundance in the {\tt TBabs} model. A constant multiplier is included in the model to account for calibration uncertainties across instruments. The constant factor is fixed at a value of 1 for \nustar{}/FPMA, while it is allowed to vary for \nustar{}/FPMB.

\begin{figure}
\includegraphics[width=0.74\columnwidth, angle=270]{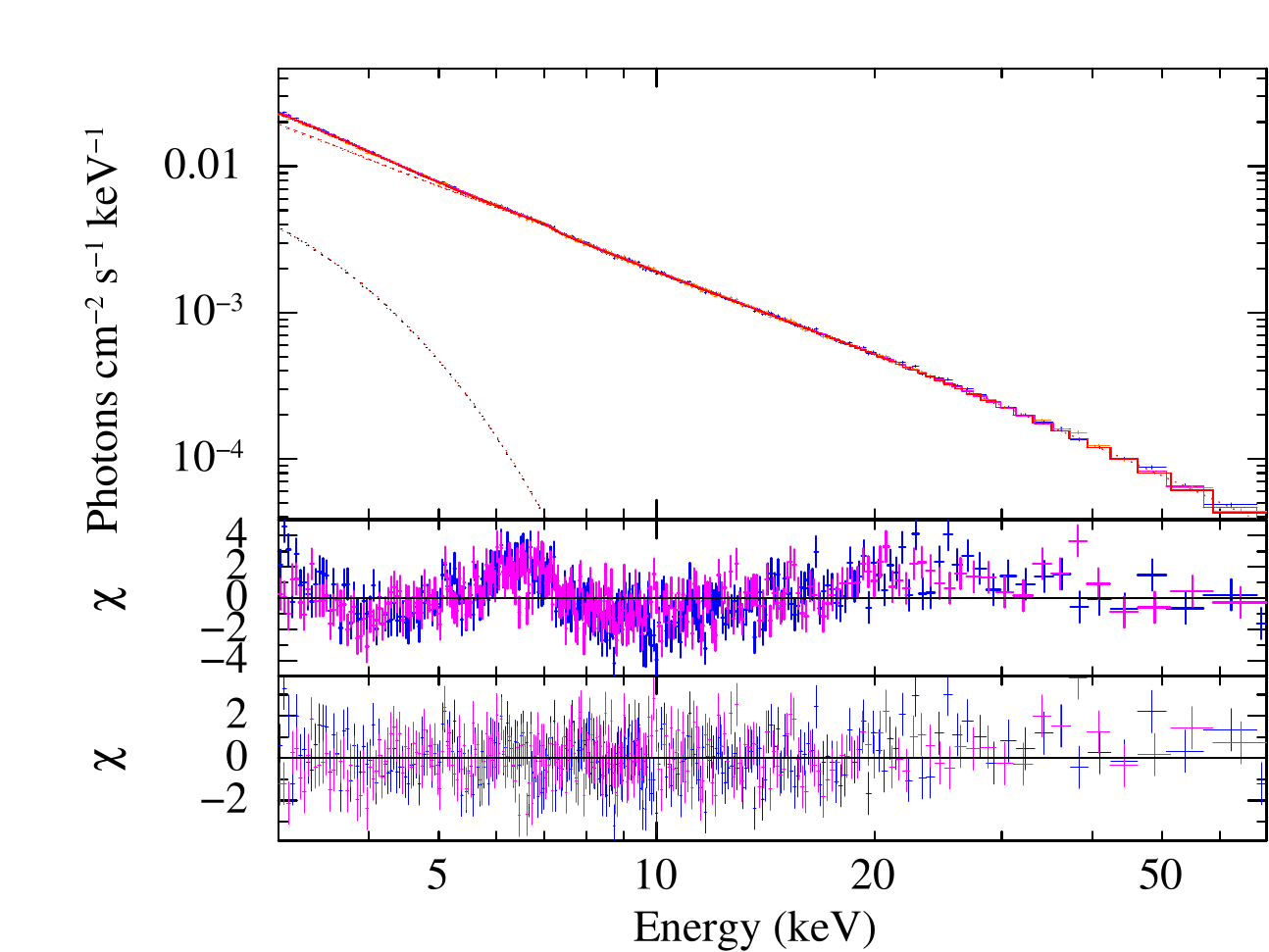}
\caption{\nustar{} spectra of \src{}, fitted with the absorbed blackbody model along with the reflection component \texttt{tbabs $\times$ (bbodyrad + relxillCp)} and corresponding residuals are shown in the top and bottom panels, respectively. The middle panel shows the residuals obtained from fitting the spectra with \texttt{tbabs $\times$ (bbodyrad + nthcomp)} model.}
    \label{fig:reflection_nustar}
\end{figure}

The single-component model (e.g., {\tt TBabs $\times$ nthComp}) provides a poor fit to the data, with a reduced chi-square of $\chi^2 \ge 2.1$ for 2301 degrees of freedom and strong residuals across the spectrum. Other single-component continuum models (e.g., {\tt TBabs $\times$ cutoffpl}) were also tested, but all yielded similarly unacceptable fits ($\chi^2 \gtrsim 2$) and left significant residuals, indicating that a single emission component is insufficient to model the spectra. Adding a soft thermal component substantially improves the fit. To describe the persistent spectra, a more sophisticated two-component model, {model M0: \tt TBabs $\times$ (bbodyrad + nthComp)}, is introduced, improving the fit compared to the single-component model. This model yields a photon index $\Gamma = 1.9$, a blackbody temperature $kT_{\rm bb} = 0.6$ keV, and an electron temperature of $\sim$53 keV. The fit statistic improves to $\chi^2 = 1.2$ for 2300 degrees of freedom (see Table~\ref{tab:fitstat2}). Despite this improvement, clear residuals remain at $\sim$6.4 keV and above $\sim$20 keV, corresponding to an iron line and a Compton reflection hump (see Fig.~\ref{fig:reflection_nustar}). The residuals, shown in the middle panel of Fig.~\ref{fig:reflection_nustar}, indicate X-ray reflection features \citep{Fa89, Fa00, Mi07}, including an iron line around 6.4 keV and a back-scattering Compton hump above $\sim$20 keV, irrespective of the choice of continuum component. These features motivate the inclusion of a reflection component, yielding a statistically acceptable fit (reduced $\chi^2 = 0.95$).  
  
To explain the X-ray reflection features, we used the self-consistent relativistic disc reflection model {\tt rexillCp} which is a member of the {\tt relxill} (v2.3: \citet{Ga14, Da14}) model family. The model {\tt relxillCp} includes the thermal Comptonization model ({\tt nthComp}) as an illuminating continuum. In contrast, the standard {\tt relxill} model uses a simple cutoff power-law ({\tt cutoffpl}) as the illuminating spectrum. We used both of these models to probe the reflection features in the persistent emission of the \nustar~ spectrum. The primary source is assumed to have an emissivity given by Index~1 (q1) and Index~2 (q2). The reflection model {\tt relxillCp} parameters are as follows: the photon index ($\Gamma$) of the illuminating radiation, inner disc radius ($R_{\mathrm{in}}$), outer disc radius ($R_{\mathrm{out}}$), disc inclination angle ($i$), dimensionless spin parameter ($a^\ast$), ionization parameter ($\log\xi$) at the surface of the disc, iron abundance ($A_{\mathrm{Fe}}$), electron temperature (kT$_e$), and disc density ($\log\text{N}$). 

The emissivity indices are fixed at q1 = q2 = 3, assuming a single emissivity. The outer radius of the accretion disc ($R_\text{out}$) is fixed at 400 gravitational radii $R_g$ ($R_g = \frac{GM}{c^2}$). The hydrogen column density is fixed at $0.21 \times 10^{22}~\mathrm{cm^{-2}}$ \citep{Salvo2019}. The spin frequency of \src~ is $\nu=401\,\mathrm{Hz}$, corresponding to a spin period $P=1/\nu=2.49\,\mathrm{ms}$. Using the relation $a_* \simeq 0.47/P(\mathrm{ms})$ \citep{Braje2000}, we obtain the spin parameter $a^\ast \approx 0.19$. Given the low spin parameter of \src~ ($a^\ast \simeq 0.19$), the ISCO depends only weakly on $a^\ast$. Adopting $a^\ast = 0$ instead of $a^\ast \approx 0.2$ changes the ISCO by $\lesssim 1\,R_g$, well within the spectral uncertainties \citep{Lu19, Ga08}, and thus does not significantly affect the inferred inner disk radius or magnetic field estimate. However, we use the spin parameter $a^\ast \approx 0.19$  during the reflection spectral modeling.

 We used both models {\tt relxill} and {\tt relxillCp} to check the consistency of the disc parameters. The X-ray reflection features in the \nustar{} persistent spectrum are modeled using the combination of models M1: {\tt TBabs $\times$ (bbodyrad + relxillCp)} and model M2: {\tt TBabs $\times$ (bbodyrad + relxill)}. The best-fit model gives a statistically acceptable fit with a reduced $\chi^2$ of $\sim$1. The best-fit \nustar{} spectra of \obj{}, along with the model M1, are shown in the top panel of Fig.~\ref{fig:reflection_nustar}, and the corresponding residuals are shown in the bottom panel. The best-fit spectral parameters for model M1 provides an inner disc radius R$_\text{in}$, and disc inclination $i$ of $2.6^{+1.8}_{-1.1}~R_{\mathrm{ISCO}}$, and $42.4^{+18.2}_{-4.4}~{}^{\circ}$. The photon index and electron temperature are found to be $2.0\pm0.02$ and $81_{-10}^{+32}$ keV, respectively. The disc density $\log \text{N} = 15.1_{-0.03}^{+1.2}~\text{cm}^{-3}$ and the disc ionization parameter is $\log \xi = 2.8^{+0.1}_{-0.3}~\mathrm{erg~cm~s}^{-1}$. The blackbody temperature and normalization of the {\tt bbodyrad} model are found to be $0.64_{-0.02}^{+0.04}$ keV and $45_{-13}^{+8}$, respectively. The corresponding apparent emitting radius of the blackbody is estimated to be $2.4_{-0.4}^{+0.2}$ km, by assuming a source distance of 3.5 kpc. The total unabsorbed flux in the range of 0.1-100 keV is $\sim 2.1 \times 10^{-9}~\mathrm{erg~cm^{-2}~s^{-1}}$, and the corresponding luminosity is estimated to be $\sim 3.1 \times 10^{36}~\mathrm{erg~s^{-1}}$ assuming a source distance of 3.5 kpc. The best-fit spectral parameters for model M2 provide an inner disc radius R$_\text{in}$, and disc inclination $i$ of $2.7^{+1.2}_{-1.1}~R_{\mathrm{ISCO}}$, and $47.4^{+5.3}_{-5.1}~{}^{\circ}$, respectively. The parameters obtained from the fitting of the data with both models M1 and M2 are given in Table~\ref{tab:fitstat2}. The accretion disc parameters obtained from two different reflection model combinations (M1 and M2) are found to be consistent (within errors). The Markov Chain Monte Carlo ({\tt MCMC}) approach is used to evaluate errors in the spectral parameters. Based on the Goodman-Weare algorithm \citep{Good2010}, the {\tt MCMC} simulation is performed on the reflection models ({\tt relxillCp}) with a chain length of 500000 and 10 walkers. The first 50000 steps are discarded, assuming they are in the `burn-in' stage.
\begin{table}
\scriptsize
	\centering
	\caption{Best-fit spectral parameters obtained by fitting the source spectra in the energy range 3--70 keV with different model combinations. Model M0 represents the spectral model \texttt{tbabs $\times$ (bbodyrad + nthComp)}. The best fits are obtained with the reflection models M1: \texttt{tbabs $\times$ (bbodyrad + relxillCp)} and M2: \texttt{tbabs $\times$ (bbodyrad + relxill)}.}
	\label{tab:fitstat2}
    \setlength{\tabcolsep}{2.7pt}   
	\begin{tabular}{lccc} 
\hline
 Parameters	&  M0 & M1& M2	\\	
\hline																			
N$_\text{H}$ ($10^{22}$ cm$^{-2}$) 	&	$0.21^c$ & $0.21^c$ & $0.21^c$  \\		
kT (keV) & $0.60\pm0.01$ & $0.64_{-0.02}^{+0.04}$ & $0.60_{-0.04}^{+0.03}$    \\
Norm$_\mathrm{bb}$	& $107_{-7.9}^{+9.5}$ & $45_{-13}^{+8}$ &	$56.2_{-13}^{+17}$\\ 
 R$_\mathrm{bb}^d$ (km)	 & $3.6_{-0.1}^{+0.2}$ &$2.4_{-0.4}^{+0.2}$ &	$2.6_{-0.3}^{+0.4}$\\
Photon index ($\Gamma$)	 & $1.90\pm0.01$ &$1.96\pm0.02$ & $2.0\pm0.03$ \\ 
kT$_{\rm e}$ (keV)  & $53.3_{-9.2}^{+17.5}$ &$81_{-10}^{+32}$ &  $-$\\ 
$E_{\mathrm{cut}}^e$ (keV)	 &- &  	$-$ &  $238_{-47}^{+27}$ \\ 
Incl (deg) 	&	- &  $42.4_{-4.4}^{+18.2}$ 	&	 $47.4_{-5.1}^{+5.3}$  \\ 
$R_{\mathrm{in}}$ ($R_{\mathrm{ISCO}})$	&	- & 	 $2.6_{-1.1}^{+1.8}$ 	&	$2.7_{-1.1}^{+1.2}$  \\
$R_{\mathrm{in}}$ ($R_{\mathrm{g}})$	 &	- &$14.0_{-5.9}^{+9.7}$	  	&	$14.6_{-5.9}^{+6.4}$  \\
$A_{\mathrm{Fe}}$	 & -&	 $0.9\pm0.3$ 	&	$1.0_{-0.2}^{+0.6}$  \\
q1=q2	 &		- & $3^c$ 	&	 $3^c$  \\ 
$\log \xi\;(\mathrm{erg~cm~s}^{-1})$ &	- &  $2.8_{-0.3}^{+0.1}$ 	&	$2.1_{-0.3}^{+0.4}$  \\
$\log\text{N}$ (cm$^{-3}$) 		& - &$15.1_{-0.03}^{+1.2}$  	&	 \\
$\mathrm{Refl}_{\mathrm{frac}}$	&	- & $0.20_{-0.03}^{+0.13}$ 	&	$0.30\pm0.05$  \\
$\mathrm {Norm}_{\mathrm{Refl}}$ ($\times10^{-3})$ 	&- &$2.40\pm0.05$ 	&	$2.7\pm0.1$  \\
\hline 
Total Flux$_{0.1-100~\mathrm{keV}}^a$ 	&	 & $2.14\pm0.01 $ 	&	$2.14\pm0.01$  \\
Luminosity$_{0.1-100~\mathrm{keV}}^b$ 	  	&	& $3.13\pm0.02$ 	& $3.13\pm0.02$ \\
\hline                   															
 $\chi^2$/dof  &	2682/2300 & 2195/2294         &  2190/2295    \\
\hline		
\multicolumn{3}{l}{$^a$ : Unabsorbed flux in the units of $10^{-9}$ \erg.}\\
\multicolumn{3}{l}{$^b$ : in the units of $10^{36}$ \lum, for a distance of 3.5 kpc.}\\
\multicolumn{3}{l}{$^c$ : Frozen parameters.}\\
\multicolumn{3}{l}{$^d$ : 
{ $R_{\rm bb} = d_{10}\sqrt{\mathrm{Norm}}$ km, apparent radius of emitting region.}}\\
\multicolumn{3}{l}{$^e$ : { $E_{\rm cut}$, high-energy cutoff of illuminating power-law continuum.}}\\ 
 \multicolumn{3}{l}{$^f$ : {bbodyrad normalization is reported in standard XSPEC units \textrm{$(R_{km}/D_{10})^2$.}}}\\
\end{tabular}
\end{table} 
\section{Discussion}
\label{dis}
We present the results of the analysis of thermonuclear X-ray bursts from \obj{} using a \nicer{} and \xmm{} EPIC-PN observations. The energy dependence of the burst profile is investigated, indicating a strong dependence of the burst profile on energy, along with a secondary peak around 15 s from the onset of the \nicer{} burst. The secondary peak is prominent in the energy ranges below 5 keV. The evolution of the hardness ratio was also investigated to find any changes in the spectral state during the burst.  The hardness ratio shows a notable evolution during the burst, reaching a minimum at the peak, indicating that the low-energy photons dominate over the high-energy photons during this phase. 
\subsubsection{Time resolved spectroscopy: burst-disc interaction}
The variable persistent emission approach is followed to describe the time-resolved burst spectra from the \nicer{} observation. However, the peak value of the $f_a$ parameter exceeded 150, which is outside the typical range and not physically meaningful. In addition, the reduced \( \chi^2 \) values obtained from the \( f_a \) modeling are statistically unacceptable, with \( \chi^2 > 2 \) for most burst segments, and the model fails to account for the large residuals below 1.8~keV. A similar behavior was reported by \citet{Bu19}. Moreover, the obtained scaling factor, $f_a$ = 159, is unrealistically high, far exceeding the typical range of 2–10 observed in previous studies \citep{Worpel2013}. For comparison, in 4U~1820-30, the photosphere expanded to over $\sim$200 km while the maximum value $f_a$ parameter was only $\sim$10 \citep{Ke18b}. In our case, the photosphere expands to a maximum of $\sim$15 km, yet $f_a$ exceeds 150, which is difficult to reconcile physically. The extremely intense burst observed with \nicer{} provides a unique opportunity to investigate the burst-disc interaction and the related emission mechanism using the latest relativistic disc reflection model {\tt relxillNS}. A detailed time-resolved spectral study of the X-ray burst is conducted to investigate the dynamic evolution of the spectral parameters. During the \nicer{} burst, the radius expansion of the photosphere is observed. By applying the color correction factor of $f_c = 1.4$ \citep{Suleimanov2011} and scaling the radius with the correction factor of $f_c^{2}/(1+z)$ (assuming a redshift of 0.3 and a source distance of 3.5 kpc), the photosphere is found to expand to a maximum of $14.8\pm0.7$ km.

At the beginning of the PRE phase, the bolometric flux increases until it reaches its maximum, close to the Eddington limit at the peak of the burst, and then remains approximately constant before gradually decreasing toward the end of the phase. During this phase, the blackbody flux remains nearly constant. This behavior can be understood from the Stefan–Boltzmann law: as the photosphere expands during PRE, the effective emitting area increases while the blackbody temperature correspondingly decreases, maintaining an approximately constant luminosity near the Eddington limit. The expansion of the photosphere also modifies the illumination geometry of the system. In particular, the enlarged photosphere increases the solid angle subtended by the accretion disk, allowing the disk to intercept a larger fraction of the burst emission. This enhanced interception of burst photons temporarily increases the reflected flux from the disk, which manifests observationally as reflection features in the burst spectra.

Spectra of Type-I X-ray bursts frequently exhibit signatures of disk reflection. These features commonly include Fe emission lines and an additional soft excess component \citep{Degenaar2013,Speicher2022,Ke18a,Bult2019,Chen2019}. During the intense bursts, the observed spectra deviate significantly from a pure blackbody emission. In the case of \src, the excess emission in low energy (< 2keV) and high energy (> 5 keV) during the burst (panel~B of Fig.~\ref{fig:reflection_nicer}), is well accounted for by the reflection model. A similar excess emission at both the low- and high-energy ends of the spectrum has also been reported in 4U~1636$-$536 \citep{Zh22}. In that study, the strong excess and the deviation from a pure blackbody spectrum were modeled using an additional reflection component (relxilNS with $R_f$ = -1) originating from the surrounding accretion disk \citep{Zh22,Speicher2022,Lu23}. This work also showed that during the first few seconds of the PRE bursts, the reflected flux can exceed the direct burst emission flux \citep{Zh22,Keek2014}, which has been interpreted in terms of the accretion disk geometry \citep{He2016}.

For an inclination of $60^{\circ}$ in 4U~1636$-$536 \citep{Pandel2008}, a thin-disk geometry predicts a reflection fraction (the ratio of the reflected flux to the illuminating blackbody flux) of about $\sim$1/3 \citep{Fujimoto1988}. \citet{Lapidus1985} claimed that the value of the reflection fraction can increase by several tens of percent due to relativistic light bending. However, the much larger reflection fraction observed (up to $\sim$6) suggests a geometrically thick disk that partially obscures the neutron star from the observer \citep{Blackman1999}. In such a scenario, the inner disk intercepts a large fraction of the burst radiation, leading to reflected emission dominating the observed spectrum.

A similar behavior has been observed during the PRE phases of thermonuclear bursts in the X-ray binary 4U~1820$-$30, where the blackbody flux peaks near the touchdown phase, while the reflection component dominates over the blackbody flux between the PRE peak and touchdown \citep{Ja24,Yu24}. This behavior is consistent with our results for the PRE burst in \src. Similar results have been reported for PRE bursts in sources such as 4U~1820$-$30 \citep{Ja24,Yu24}, 4U~1636$-$536 \citep{Zh22,Keek2014}, and 4U~1702$-$429 \citep{4U1702}. In contrast, non-PRE bursts, such as those observed from Aql~X-1 \citep{Ma25} and from \src\ (\xmm\ observation; present work), have been analyzed using a similar \texttt{relxillNS} setup but do not show the reflection component dominating over the blackbody flux. This reflection-dominated behavior, therefore, appears to be specific to PRE bursts and is likely associated with changes in the illumination geometry induced by the expanding neutron star photosphere.

 During the PRE phase, the neutron star photosphere expands significantly, causing the emitted radiation to become strongly anisotropic as in the present case of \src. As a result, the observer receives only a fraction of the total emitted flux due to angular redistribution and possible partial obscuration by the accretion disk. At the same time, the expanded photosphere increases the solid angle subtended by the accretion disk, allowing it to intercept a larger fraction of the burst emission. Consequently, the reflection fraction can temporarily exceed unity near the peak of the PRE burst without violating energy conservation, consistent with theoretical expectations for strong PRE bursts \citep{Ba04,Degenaar2016}. Depending on the disk geometry, particularly if the disk is vertically extended, the reflection fraction can exceed unity \citep{He2016}. In such cases, the reflected flux may dominate over the directly observed blackbody emission during the PRE phase as observed in the PRE burst of \src~ with \nicer. This situation can be illustrated by the analogy of a candle in a teacup: even when the flame is not directly visible from certain viewing angles, its light remains observable after scattering off the inner walls of the cup \citep{Ga21}.

In addition, an emission line at 1 keV is observed in the \nicer{} burst spectra. Detection of the 1 keV emission line has been reported during X-ray bursts in several other sources, including 4U~1820-30 \citep{Degenaar2013, Ja25}, IGR~J17062-6143 \citep{Degenaar2013}, and HETE J1900.1-2455 \citep{Papitto2013}. This emission line may have originated from the Fe L-band transition because of the reprocessing of burst photons by cold gas in the accretion disc. It is also possible that the 1 keV emission line may originate from the Ly$\alpha$ transition of Ne~X. Earlier, it was reported that ultra-compact X-ray binaries exhibit an overabundance of Ne \citep{Degenaar2013}. The centroid energy of the emission line (E) is $\sim$1.02 keV, and the corresponding average width ($\sigma$) is 0.082 keV (FWHM = 0.192 keV) during the peak of the burst. The velocity of broadening ($\frac{v}{c}=\frac{\text{FWHM}}{\text{E}}$) can be estimated to be $v\sim0.19 c$. Assuming that the material is in a Keplerian orbit, the radial distance is estimated to be ${\textrm r=\frac{Gm}{v^2}= 57}$ km from the neutron star (assuming a typical NS of radius and mass of 10 km and 1.4$~M_\odot$, respectively). The velocity estimate $v/c = \mathrm{FWHM}/E \sim 0.19$ does not account for the fact that the full width corresponds to twice the Keplerian velocity, nor for the inclination of the system, which reduces the line-of-sight component. Correcting for these effects, the Keplerian velocity is $v_{\rm Kep} \approx 0.12\,c$, which increases the radial distance from 57 km to $\sim 135$ km for an inclination of $50^\circ$. 

Further, the characteristics of the line are investigated by estimating the line flux and its correlation with the flux of the reflection and blackbody components. The 1 keV emission line flux is found to be strongly correlated with the flux of the reflection component. The strong correlation between these two components indicates that they may originate from a common physical process during reprocessing of burst emission through the accretion disc. In the {\tt relxillNS} approach, the reflection spectrum is produced when the photoionized plasma reprocesses thermal photons from the neutron star surface or boundary layer that illuminate the accretion disc. The blended Fe L-shell and Ne~X Ly$\alpha$ transitions in a moderately ionized disc are most likely the source of the 1 keV emission line. The strong correlation between the 1 keV line flux and the reflection component flux indicates that both vary in response to changes in the illumination and ionization state of the disc as the burst evolves.

A detailed time-resolved spectral study is performed for the thermonuclear burst observed with \xmm{} EPIC-PN. The burst spectra are modeled using an absorbed blackbody model, which shows an excess emission in the soft X-ray range. This is further investigated using the variable-persistent-emission method and the {\tt relxillNS} disc reflection model. Several possible explanations for the excess emission include the atmospheric effect of NS \citep{Ozel2013} and Poynting–Robertson drag \citep{Walker1992, Worpel2013}. Alternatively, the accretion disc may interact with a fraction of the burst photons, and these photons may then be reflected from the disc. We found that a more realistic disc reflection model, {\tt relxillNS}, can describe the burst spectra and provides a self-consistent and physically realistic explanation. This approach was adopted to model the time-resolved burst spectra for several similar types of sources such as 4U~1636-536 \citep{Zh22}, Aql~X-1 \citep{Ma25}, 4U~1730-22 \citep{Lu23}, SAX~1808.4-3658 \citep{Bu21}, 4U~1820--30 \citep{Ke18a, Ja24}, 4U~1702-729 \citep{4U1702}, and SRGA~J144459.2-604207 \citep{SRGA25}. 
\subsubsection{Thermonuclear ignition regime}
The mass accretion rate is estimated from pre-burst spectral fitting and is further utilized to investigate the thermonuclear bursting regime. The local accretion rate to the NS can be expressed as \citet{Ga08} 

\begin{align}
\dot{m} =\; & \frac{L (1+z)}{4\pi R^2 (GM/R)} \nonumber \\
=\; & 6.7 \times 10^{3} 
\left( \frac{F_{b}}{10^{-9}~\mathrm{erg~cm^{-2}~s^{-1}}} \right) 
\left( \frac{d}{10~\mathrm{kpc}} \right)^2 
\left( \frac{1+z}{1.31} \right) \nonumber \\
& \times \left( \frac{R}{10~\mathrm{km}} \right)^{-1} 
\left( \frac{M}{1.4~M_\odot} \right)^{-1} 
~\mathrm{g~cm^{-2}~s^{-1}}
\label{eqn1}
\end{align}

where $M$ and $R$ are the mass and radius of NS, $d$ is the distance to NS, and $F_b$ is the persistent bolometric flux of \obj{} (Table~\ref{tab:tab_preburst}). The mass accretion rate is estimated from the \nicer{} and \xmm{} EPIC-PN preburst flux by assuming a typical NS of mass 1.4$~M_\odot$, radius 10 km, gravitational redshift ($z$) of 0.3, and the source distance of 3.5 kpc. The bolometric preburst flux during the \nicer{} and \xmm{} observations was in the range of 5.0--5.2 $\times 10^{-10}$ \flux. Using these values in Equation~\ref{eqn1}, the corresponding mass accretion rate is estimated to be $0.041-0.043~\times~10^{4}$ gm~cm$^{-2}$~s$^{-1}$. The details related to the mass accretion rate are summarized in Table~\ref{tab:tab_preburst}. The local Eddington mass accretion rate for a typical NS with radius 10 km is $\dot{m}_{\mathrm{Edd}} = 8.8~\times~10^4~\mathrm{g~cm^{-2}~s^{-1}}$. The local accretion rate is related to the Eddington rate as $\dot{m} = (0.0047-0.0049)~\dot{m}_{\mathrm{Edd}}$, indicating an unstable ignition of hydrogen, which triggered a type-I X-ray burst in a hydrogen-rich environment \citep{Ga08}. 

 At this level of low mass accretion rate, hydrogen burns unstably into helium. Although at this accretion rate the burst ignition is expected to be helium-triggered, residual hydrogen may survive until ignition and contribute to the burst energetics, and it is also possible that a large He layer may build up, producing an extremely high-energetic, long X-ray burst \citep{Peng2007}. The PRE burst duration is comparatively longer (>20 s) than that of a pure helium burst ($\sim$10 s), which indicates that the bursts are possibly powered by a helium-rich mixed H/He fuel composition \citep{Ga08}. The observed burst duration (>20 s), together with the presence of photospheric radius expansion, is therefore inconsistent with a classical pure-helium burst and instead suggests ignition in a helium-rich but mixed H/He fuel layer.
\subsubsection{X-ray reflection in the \nustar{} persistent spectrum}
The X-ray reflection features are also investigated from the persistent emission using the \nustar{} observation of \obj. The relativistic disc reflection model \relxill{}/\relxillCp{} is used to investigate these features further. The X-ray reflection features are used to probe disc geometry and properties with modern high-resolution telescopes such as \xmm{}, \nustar{}, and {\it INTEGRAL} in several sources \citep{Pa10, Pa13a, Pa13b, Pa16, 4U1702, Ma25, Lu19, Ch22, Ludlam2024, Malacaria2025}. The \nustar{} spectra of \obj{} show an iron line at around 6.4 keV and a Compton hump at around 20 keV, indicating a reflection feature. The persistent spectrum is modeled with an absorbed blackbody and a relativistic disc reflection component. Reflection is described using {\tt relxill} and {\tt relxillCp}, which assume a cutoff power-law ({\tt cutoffpl}) and a thermal Comptonization ({\tt nthComp}) illuminating continuum, respectively, included self-consistently in the models. The best-fit results provide an inclination angle of $38^\circ-60^\circ$ and an inner disc radius of $14_{-5.9}^{+9.7}$ $R_g$. The accretion disc is found to be moderately ionized with the ionization parameter $\log\xi$ of $2.8_{-0.3}^{+0.1}$ erg cm s$^{-1}$. Previously, the X-ray reflection feature was investigated by \citep{Salvo2019} during the 2015 outburst using \xmm{} and \nustar{}. They reported an inner disc radius of $\sim$12 $R_g$ and an inclination angle of more than 50$^\circ{}$. Our results from reflection spectroscopy during the 2022 \nustar{} observation closely match the earlier findings. However, the flux was higher during the 2015 observation than during the 2022 observation. A previous study using optical observations reported an inclination angle of 36$^{\circ}$-67$^{\circ}$ \citep{Deloye2008}.

In the case of the NS LMXB, the low magnetic field allows the accretion disc to extend close to the magnetosphere boundary, and it is expected that the disc will be truncated at a moderate radius \citep{Ca09}. The magnetic dipole moment and magnetic field of \src~ are estimated by assuming that the accretion disc is truncated at the magnetosphere boundary. The magnetic dipole moment ($\mu$) and the magnetic field can be estimated by using the equation from \citet{Ib09}, 

\begin{align}
\mu =\; & 3.5 \times 10^{23} \, x^{7/4} \, k_A^{-7/4} 
\left( \frac{M}{1.4\, M_\odot} \right)^2 \nonumber \\
& \times \left( \frac{f_{\rm ang}}{\eta} \frac{F_b}{10^{-9}\, \mathrm{erg\, cm^{-2}\, s^{-1}}} \right)^{1/2}
\left( \frac{D}{3.5\, \mathrm{kpc}} \right)
\label{eqn2}
\end{align}

where, $M$ is the mass of the NS, $k_A$ is the geometric coefficient, $\eta$ is the accretion efficiency, $\rm f_{ang}$ is the anisotropy correction factor, and $F_b$ is the unabsorbed bolometric flux in the 0.1--100 keV range. The scaling factor $x$ can be estimated from $\text R_{\rm in} = \frac{xGM}{c^2}$, and the value of the factor is $\sim$15. Assuming $f_{\rm ang} = 1$, $\eta = 0.1$, and $k_A = 1$ \citep{Ca09} in Equation~\ref{eqn2}, the magnetic dipole moment of the NS can be estimated to be $\mu = 1.85 \times 10^{26}$ G~cm$^3$ for the flux of $\text F_b \sim 2.14~\times~10^{-9}$ erg~cm$^{-2}$~s$^{-1}$. The magnetic field strength at the poles of the NS of radius 10 km is estimated to be $ B~\simeq3.7~\times~10^8$ G. The estimated magnetic field of \src~ is in the typical range of NS LMXBs \citep{Burderi2003, Salvo2019,Mu15, Ca09, Lu16}.
\section{Conclusion}
\label{con}
We investigated the disc reflection feature and the burst-disc interaction and related mechanism for the most energetic X-ray burst from \src~ observed with \nicer{}. The burst profile is found to be energy dependent. The burst is very intense in the soft X-ray range, with a prominent secondary peak. The hardness ratio is found to evolve significantly during the X-ray burst observed with \nicer{}. A time-resolved spectral study is conducted to probe the dynamic evolution of the spectral parameters, indicating a PRE burst from \obj{} with a maximum photospheric expansion of $\sim$15 km, as observed with \nicer{}. The relativistic disc reflection model {\tt relxillNS} is utilized to explain the deviation of the burst emission from a pure blackbody model during the burst. In addition, an emission feature at 1 keV is observed, potentially originating from the accretion disc due to the Ne~ or Fe~L band transition. The velocity broadening corresponds to a Keplerian velocity of $v_{\rm Kep} \simeq 0.12\,c$ after correcting for inclination and the factor of two between the full line width and orbital velocity, implying a radial distance of $\sim 135$ km from the neutron star. The emission line flux is found to be strongly correlated with the flux of the reflection component. 

We also carried out a detailed time-resolved spectral analysis of a thermonuclear burst of \obj{} using \xmm{} EPIC-PN to probe the evolution of the burst spectral parameters and emission mechanism. The burst emission can be well modeled by an absorbed blackbody, and the excess emission is further explained using the variable persistent emission approach and disc reflection modeling. The mass accretion rate is estimated to probe the bursting regime. Based on the estimated mass accretion rate, the PRE X-ray burst is powered by He or a mixed H/He fuel. Moreover, the X-ray reflection features are also investigated in persistent emission using a \nustar{} observation, which provides a disc inclination and inner disc radius of $ 38^\circ-60^\circ$ and $14_{-5.9}^{+9.7}$ $R_g$, respectively. The magnetic field at the poles of the neutron star is estimated to be $\simeq3.7 \times 10^8$ G by assuming the disc is truncated at the magnetosphere boundary.
\section*{Acknowledgements}
We thank the anonymous reviewer for useful comments, which helped to improve the manuscript. The research work at the Physical Research Laboratory, Ahmedabad, is funded by the Department of Space, Government of India. This research has made use of data obtained with \nustar{}, a project led by Caltech, funded by NASA, and managed by NASA/JPL, and has utilized the {\tt NUSTARDAS} software package, jointly developed by the ASDC (Italy) and Caltech (USA). We acknowledge the use of public data from the \nustar{}, \nicer{}, and \xmm{} data archives.
\section*{Data Availability}
The data used for this article are publicly available in the High Energy Astrophysics Science Archive Research Centre (HEASARC).
\bibliographystyle{mnras}
\bibliography{SAX_MNRAS}
\pagestyle{plain}
\bsp	
\label{lastpage}
\end{document}